# A Dataset of the Representatives Elected in France During the Fifth Republic

Noémie Févrat[1], Vincent Labatut[2], Émilie Volpi[3], Guillaume Marrel[1]

[1] JPEG UPR 3788, Avignon Université, Avignon, France.
[2] LIA UPR 4128, Avignon Université, Avignon, France.
[3] Agorantic FR 3621, Avignon, France.
`{firstname.lastname}@univ-avignon.fr`

**Abstract:** The electoral system is a cornerstone of democracy, shaping the structure of political competition, representation, and accountability. In the case of France, it is difficult to access data describing elected representatives, though, as they are scattered across a number of sources, including public institutions, but also academic and individual efforts. This article presents a unified relational database that aims at tackling this issue by gathering information regarding representatives elected in France over the whole Fifth Republic (1958–present). This database constitutes an unprecedented resource for analyzing the evolution of political representation in France, exploring trends in party system dynamics, gender equality, and the professionalization of politics. By providing a longitudinal view of French elected representatives, the database facilitates research on the institutional stability of the Fifth Republic, offering insights into the factors of political change.

## 1.   VALUE OF THE DATA

- The *Base de données Révisée des Élu·es de France*[1] (BRÉF) is a biographical and historical database that aims to list, document and verify exhaustively, for scientific purposes, all individuals who have held a representative political office following an election throughout France, in local or national assemblies, since the start of the Fifth Republic in 1958.
- The BRÉF is intended to eventually compile all the databases on French elected politicians that are scattered across History and Political Science research teams. It is valuable because it aims to be exhaustive, it cross-references data and can be queried by name of individual, by type of mandate or by area, and each search can be located and dated.
- Our database is of interest to all researchers working on the sociology of political staff in representative democracies, identity, origins, recruitment inequalities in terms of gender, age and social origins, inheritance, entry into politics, careers and upward or downward trajectories [1], political longevity [2], political professionalization dynamics [3], the anchoring and entrenchment of elected representatives, electoral failure and party turnover, and the political history of any area of France. The BRÉF is also of interest to other Social Science researchers working on electoral behavior and local public policy.
- In addition, it can be useful in the field of Economics, and more generally it constitutes a relevant benchmarking dataset for Computer Science algorithms related to Data Mining. For this purpose, the data can be used under the form of partial extractions from the database, integrated into other databases and enriched, or connected to existing databases [4].

---

[1] Revised Database of Representatives Elected in France

## 2. BACKGROUND

The BRÉF was developed as part of Noémie Févrat's PhD work on political longevity and term limits in France [2]. One objective of this research, conducted within the field of Political Science, was to study the so-called *political trajectories* of representatives elected in France under the Fifth Republic, i.e., the succession of positions held throughout their careers. For this purpose, we leveraged various tools based on Sequence Analysis [5], which required both good geographic coverage and significant historical depth.

The most suitable available database that met these constraints was maintained by the French Ministry of the Interior. However, it was designed for different purposes, in particular monitoring the legality of candidacies in elections, and was not entirely appropriate or reliable for our goals. The BRÉF was created to fill this gap by correcting and supplementing the original data using additional sources. During its development, many researchers reached out to express their interest in this resource. Now that the BRÉF has achieved an acceptable level of quality, particularly regarding completeness and reliability, we wish to share it with the research community.

## 3. DATA DESCRIPTION

In this section, we first introduce the general architecture of the database (Section 3.1) before detailing individually each one of its constituting data tables (Section 3.2). We then review its integrity constraints (Section 3.3), and finally provide some statistics to describe its content (Section 3.4). The BRÉF is available under three forms [6]: first, as a collection of CSV files, each one corresponding to a data table; second, as an SQL dump; and third, as a binary dump of our PostgreSQL database, which can be imported in a compatible system.

### 3.1 Architecture of the database

The BRÉF was designed to store the data describing elected representatives, the terms they served, and the areas they represented, whose population elected them. Its architecture is presented in Fig.1. It allows for some flexibility, a feature which is required in order to adapt to the frequent changes of the rules and laws that control elections in France. The database is organized around three main tables (`Mandate`, `Individual` and `Area`), which describe the central objects in our modeling. Each term (table `Mandate`) is associated to a specific representative (table `Individual`) which is elected in a specific administrative subdivision (table `Area`). Terms correspond to different types of positions (field `TypeMandate` in table `Mandate`), and the representative may fulfill a specific function during their term (table `Function`). At the time of their election, a representative can declare a professional occupation (table `Profession`) and a political orientation (table `PoliticalNuance`). Finally, the areas are linked by various relations of inclusion (table `Inclusion`).

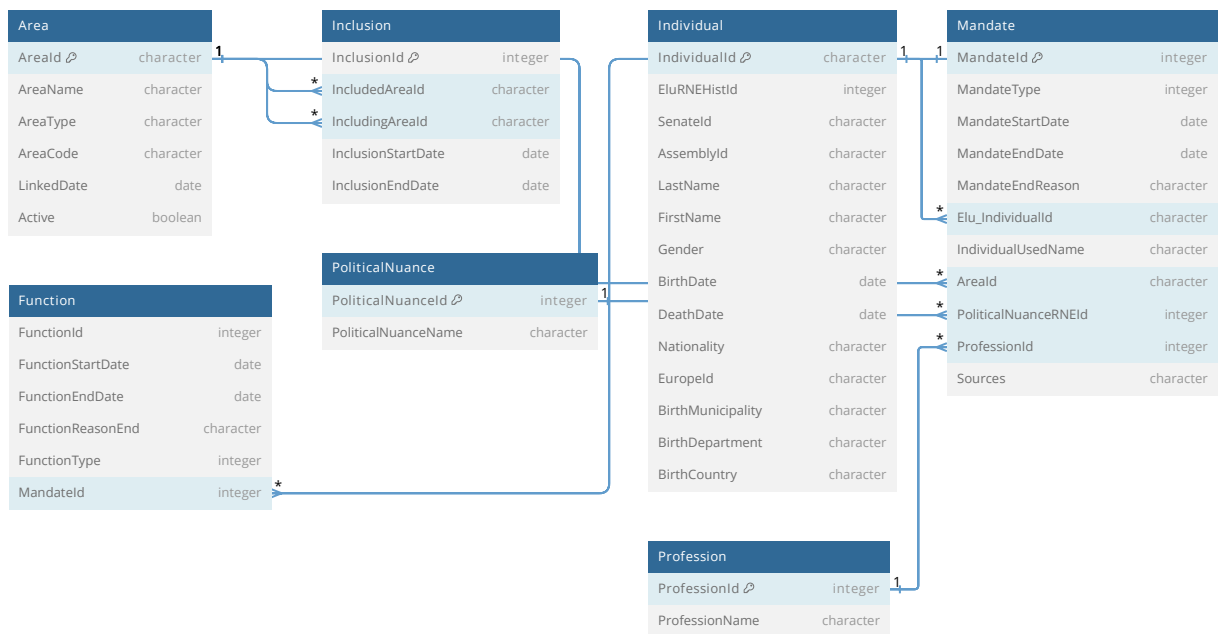

**Fig. 1.** Diagram of the data model.

## 3.2 Description of the data tables

Table 1 shows the fields of the data table `Individual`. Each one of its entries describes an individual of the database, i.e. an elected representative. A representative has a unique ID (field `IndividualId`), which is internal to the database. It takes the form `XXX_ddddddd`, where `XXX` denotes the data source in which the representative was first encountered, followed by an incremental number without specific meaning. The possible values of `XXX` are:

- `ASN`: National Assembly database;
- `EUR`: European Parliament database;
- `RNE`: French registry of elected representative;
- `SEN`: Senate database;
- `WIK`: Information found on Wikipedia.

A representative may also have several external IDs coming from the original data sources, and stored in fields, for reference: RNE (field `RNEHistId`), Senate database (field `SenateId`), National Assembly database (field `AssemblyId`), European Parliament website (field `EuropeId`).

Certain fields are only present in certain sources, and/or are optional at the time of entering the information. For this reason, they are relatively sparse. This is particularly the case of the individuals' nationality (field `Nationality`) and place of birth (fields `BirthMunicipality`, `BirthDepartment`, and `BirthCountry`).

**Table 1.** List of the fields in the data table `Individual`.

| Field | Type | Description |
|---|---|---|
| `IndividualId` | varchar | Primary key that uniquely identifies each elected representative. |
| `RNEHistId` | integer | ID of the individual in the RNE database. |
| `SenateId` | varchar | ID of the individual in the Senate database. |
| `AssemblyId` | varchar | ID of the individual in the National Assembly database. |
| `EuropeId` | varchar | ID of the individual in the European Parliament database. |
| `LastName` | varchar | Family name of the individual. |
| `FirstName` | varchar | First name of the individual. |
| `Gender` | char(1) | Declared gender: `M` for male and `F` for female. |
| `BirthDate` | date | Date of birth. |
| `DeathDate` | date | Date of death, empty if the individual is alive. |
| `Nationality` | varchar | Nationally of the individual, written in full. |
| `BirthMunicipality` | varchar | Municipality of birth of the individual, written in full. |
| `BirthDepartment` | varchar | Department of birth of the individual, written in full. |
| `BirthCountry` | varchar | Country of birth of the individual, written in full. |

Table 2 provides the list of the fields from data table `Mandate`. Each one of its entries describes a term, i.e. the fact that an individual was elected to a certain position. As a consequence, each term is linked to an individual (field `IndividualId`), and to an administrative subdivision (field `AreaId`) corresponding to the term constituency. All types of terms are stored in this table, and can be distinguished through the `MandateType` field, whose value is one of the following:

- Municipal Councilor;
- Intercommunal Councilor;
- Departmental Councilor;
- Regional Councilor;
- Member of the National Assembly;
- Senator;
- President of the Republic;
- Member of the European Parliament.

This table contains certain representative-related information that is contextual to the concerned election: the last name of the individual (field `IndividualUsedName`) and their declared professional occupation (field `ProfessionId`) at the time of the election, as well as the political orientation declared by the individual for this election (field `PoliticalNuanceRNEId`). The occupation refers to an entry in table `Profession`, and the political orientation refers to an entry in table `PoliticalNuance`.

The RNE often provides a reason regarding why a representative stopped holding a given position (field `MandateEndReason`), e.g. regular end of term, voluntary resignation, forced resignation after a justice decision, cancellation of the election, death of the representative, nomination to a governmental position such as minister, etc.

Finally, field `Sources` helps keeping track of the original data source that provided the description of the term. If there is more than one source, they are listed within the same field, and separated by a semicolon. Ex: `RNE` (single source), `RNE;Wikipedia` (two sources).

**Table 2.** List of the fields in the data table `Mandate`.

| Field | Type | Description |
| --- | --- | --- |
| `MandateId` | integer | Primary key taking the form of an auto-increment integer. |
| `MandateType` | varchar | Nature of the term served by the representative. |
| `MandateStartDate` | date | Date on which the term began. |
| `MandateEndDate` | date | Date on which the term ended. |
| `MandateEndReason` | varchar | Reasons why the term ended. |
| `IndividualId` | varchar | Foreign key indicating which representative in table `Individual` served the term. |
| `IndividualUsedName` | varchar | Family name preferred by the representative at the time of the election. |
| `AreaId` | varchar | Foreign key indicating which administrative subdivision in table `Area` constitutes the term constituency. |
| `PoliticalNuanceRNEId` | integer | Foreign key indicating which political orientation in table `PoliticalNuance` was declared by the representative. |
| `ProfessionId` | varchar | Foreign key indicating which occupation in table `Profession` was declared by the representative. |
| `Sources` | varchar | Source(s) of the data describing the term. |

Table 3 lists the fields of data table `Function`. Each one of its entries describes a function, i.e. the fact that an individual was elected or nominated to a specific position *in relation to their mandate*. As a consequence, each function is linked to a term (field `MandateId`). The type of the function (field `FunctionType`) can be for instance `Adjoint au maire` (deputy mayor) for a municipal councilor, or `Président de l'Assemblée` (president of the National Assembly) for a member of the National Assembly.

Like for the terms, the RNE often provides the reason for the end of a function. The possible options are:

- `AU`: other reason;
- `DC`: death;
- `DO`: forced resignation (can be due to a justice decision);
- `DV`: voluntary resignation;
- `FM`: regular end of function.

**Table 3.** List of the fields in data table `Function`.

| Field | Type | Description |
| --- | --- | --- |
| `FunctionId` | integer | Primary key taking the form of an auto-increment integer. |
| `FunctionType` | varchar | Nature of the function held by the representative. |
| `FunctionStartDate` | date | Date on which the function began. |
| `FunctionEndDate` | date | Date on which the function ended. |
| `FunctionEndReason` | varchar | Reason why the function ended. |
| `MandateId` | integer | Foreign key indicating which entry in table `Mandate` hosts the function. |

Table 4 contains the fields of data table `Area`, which represents the different administrative areas as well as the corresponding institutions linked to the terms. All administrative levels of France are represented, and identified through field `AreaType`, which can be one of the following:

- *Commune*: municipality;
- *Collectivité territoriale*: territorial community –only used for Corsica;
- *Établissement Public de Coopération Intercommunale*[2] *(EPCI)*: intercommunal structure, i.e. gathering of several municipalities;
- *Canton*: subdivision of a department, used as constituency for departmental elections;
- *Circonscription législative*: legislative constituency, can be a group of several small contiguous municipalities, a single municipality, or a part of a large municipality;
- *Département*: created after the French Revolution and revised several times since then, now a subdivision of a region;
- *Région*: largest subdivision, created in 1982 and revised in 2015;
- *Pays*: the whole country;
- *Circonscription Européenne*: European constituency, can be a group of regions or the whole country, depending on the considered election date.

The unique identifier of the area (field `AreaId`) is a string that depends on its type:

---

[2] Public intercommunal cooperation establishment.

- Department or territorial community: `yyyy_cc` with `yyyy` the year of creation of the department, and `cc` the original administrative code of the department (a number for metropolitan France, and letters for overseas areas);
- Canton: `cantonid-cantoncode`, where these codes come from the original RNE dataset;
- EPCI: *Système d'identification du répertoire des entreprises*[3] (SIREN) code, a national number issued by the French register of establishments and facilities;
- Other types: auto-increment integer.

Like for the representatives in table `Individual`, certain areas also have external IDs originating from other databases. These are generally issued by the *Institut National de la Statistique et des Etudes Economiques*[4] (INSEE). These IDs are stored in field `AreaCode`:

- Municipality: five-digit INSEE code `DDCCC`, where `DD` is the department code, and `CCC` is the city code;
- Department: original code of the department (a number for metropolitan France, and letters for overseas areas);
- Canton: code from the RNE database.

Two fields are designed to provide some context to the area. First, `Active` indicates whether the area still exists at the date of publication of the BRÉF, or if it is an obsolete administrative subdivision now disappeared. Second, `LinkedDate` contains the date of creation of the administrative subdivision if it is active, or its date of dissolution if it is not active anymore. Finally, the hierarchical dependencies between areas of different levels are modeled in table `Inclusion`.

**Table 4.** List of the fields in the data table `Area`.

| Field | Type | Description |
| --- | --- | --- |
| AreaId | varchar | Primary key that uniquely identifies each area (see text). |
| AreaName | varchar | Name of the area. |
| AreaType | varchar | Nature of the area (see text). |
| AreaCode | varchar | Code of the area, originating from some external typology. |
| Active | boolean | `true` if the area currently exists, and `false` if it existed in the past, but not anymore. |
| LinkedDate | date | Date of creation of an active area, or of dissolution of an inactive one. |

Table 5 focuses on the fields that constitute data table `Inclusion`. It models the hierarchical dependencies between the objects of table `Area`, e.g. a certain municipality belonging to a given department. It also represents their evolution over time, e.g. a municipality leaving some intercommunal structure for another.

Each entry in this table refers to two existing areas, one (field `IncludingAreaId`) larger than the other (field `IncludedAreaId`). In our previous example, the former would be the intercommunal structure or the department, and the latter would be the municipality. One compulsory date specifies when the inclusion relationship began (field `InclusionStartDate`), and one optional date can be used to indicate when it ended (field `InclusionEndDate`), when necessary.

---

[3] Business registry identification system
[4] French national institute of statistics and economic studies.

**Table 5.** List of the fields in the data table `Inclusion`.

| Field | Type | Description |
| --- | --- | --- |
| `InclusionId` | integer | Primary key taking the form of an auto-increment integer. |
| `IncludedAreaId` | varchar | Foreign key indicating which entry in table `Area` is the smallest in the inclusion relationship. |
| `IncludingAreaId` | varchar | Foreign key indicating which entry in table `Area` is the largest in the inclusion relationship. |
| `InclusionStartDate` | date | Date on which the relation of inclusion was created. |
| `InclusionEndDate` | date | Date on which the relation of inclusion became obsolete. |

Table 6 shows the fields of data table `PoliticalNuance`, which contains all the different political nuances one can associate to a representative for a specific election. Political nuances are categories defined by the French Ministry of the Interior. Some correspond to specific parties, usually important ones, e.g. `LR` (Republican Party), `SOC` (Socialist Party), `COM` (Communist Party). Other categories are used to refer collectively to groups of smaller similar parties or individual representatives, e.g. `DVD` (various minor right-wing parties and representatives), `REG` (regionalist parties and representatives).

**Table 6.** List of the fields in the data table `PoliticalNuance`.

| Field | Type | Description |
| --- | --- | --- |
| `PoliticalNuanceId` | integer | Primary key taking the form of an auto-increment integer. |
| `PoliticalNuanceName` | varchar | Label describing a political orientation. |

Table 7 displays the fields of data table `Profession`, which contains all the professions referred to in the `Individual` table. Professions are described in free text fields in the original data sources, as they are generally declared by the representative for each term they serve.

**Table 7.** List of the fields in the data table `Profession`.

| Field | Type | Description |
| --- | --- | --- |
| `ProfessionId` | integer | Primary key taking the form of an auto-increment integer. |
| `ProfessionName` | varchar | Label of the profession declared by the representatives at the beginning of their term. |

## 3.3 Integrity constraints

An individual can hold no function at all, a single function, or several functions, but only within the temporal bounds of a term. As a consequence, each entry of table `Function` must be connected to exactly one entry of table `Mandate`; whereas one entry from table `Mandate` can be connected to zero, one, or more from table `Function`. A representative can serve an EPCI term only if they simultaneously serve a municipal term.

The region level is the largest administrative division in France. Regions were created in 1982 and underwent a major revision in 2015. A department is fully included in a region. The department level is the stablest administrative division in France: their borders, names and code have not changed much since the French Revolution, except for very specific situations, in particular around Paris. Every area can be linked directly or indirectly to a department. There is often a single European constituency corresponding to the whole country, but there punctually were several European constituencies, corresponding to groups of regions.

A municipality is fully included in a department. A canton belongs to only one department, and a department contains several cantons. An EPCI is an intercommunal structure that gathers several municipalities, one of them being its capital. The EPCI can extend over several departments and even regions, but it is attached to the department of its capital. A municipality can be included in only one EPCI, but can also be independent, and even move from one EPCI to another. Finally, legislative constituencies are fully contained in departments. They can be a group of municipalities, a whole municipality, or part of a municipality.

## 3.4 Descriptive statistics

We provide some statistics describing the data shared online [6], i.e. after they underwent the processing described in Section 4. We first consider temporal coverage. Fig.2 shows the evolution over time of the number of terms described in the database, for each type of elected position (except that of President of the Republic, as there is always only one person holding this position at any time). For the sake of comparison, we show in red the counts reported for the raw RNE data, and in green those corresponding to the second version of the BRÉF, which results from a number of corrections and modifications (cf. Section 4). As a consequence, the zones rendered in brown correspond to terms present in *both* databases. The dashed black line at the top of each plot materializes the expected number of terms, as specified by regulations or laws (it evolves through time, depending on these laws and regulations), whereas the vertical dotted black lines show the election dates. In the case of departmental seats (Figure 2.a), only half of the councilors were renewed at each election, until 2015. There were consequently two groups of seats, numbered 1 and 2 at the top of the figure, whereas the underscore denotes the switch to a full renewal of the councilors. The situation is similar for Senators (Figure 2.g), except each election initially concerned a third of the seats (hence the three groups a, b, and c), before switching to half the seats in 2011 (hence the two groups 1 and 2).

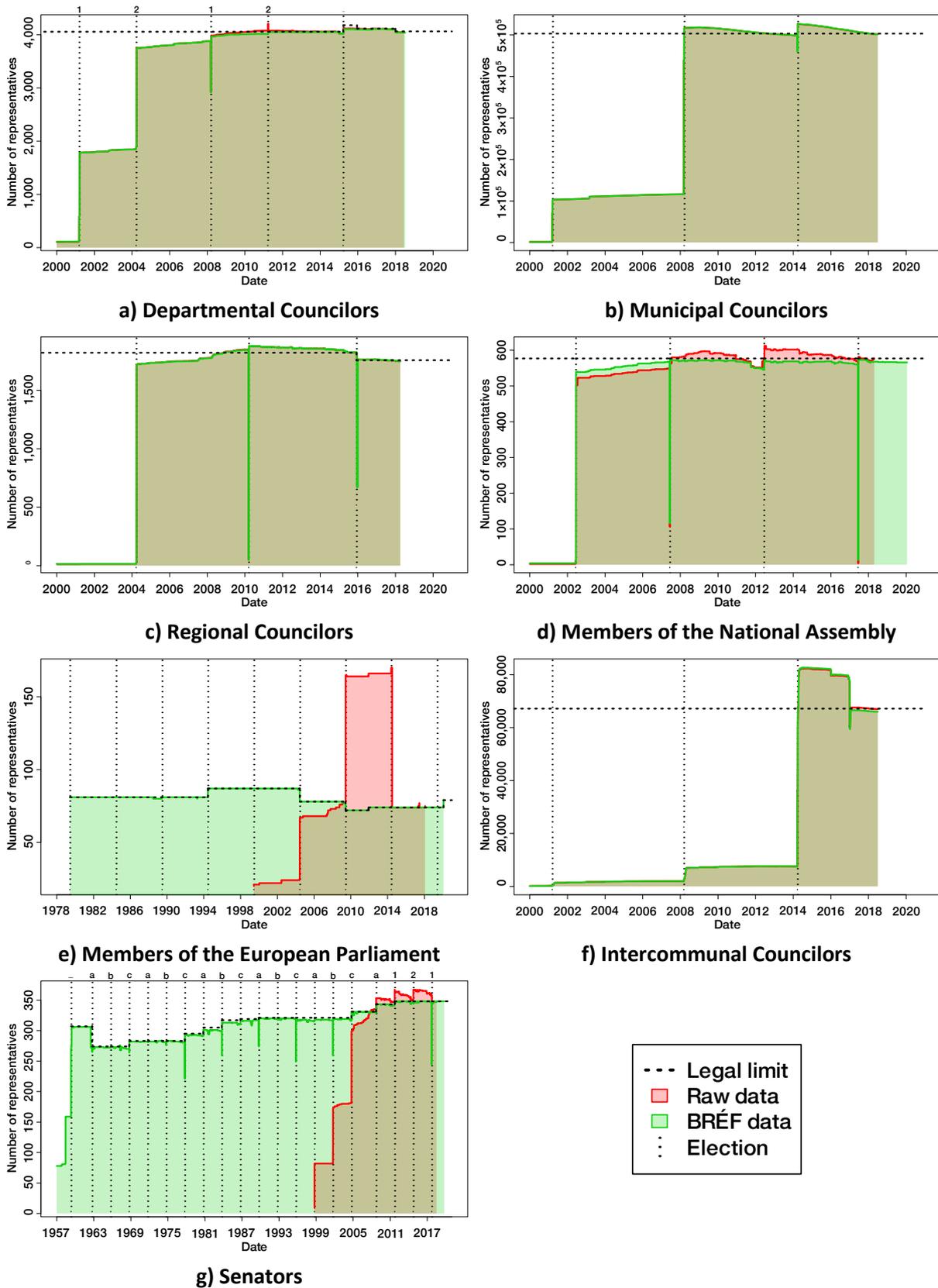

**Fig. 2.** Evolution of the number of terms over time, in the raw data (red) and in the BRÉF. Each plot corresponds to a specific type of position. The solid black lines represent the expected numbers of representatives according to laws and regulations. The vertical dotted black lines show the election dates.

The covered time range depends on the type of position, as certain were created long after the beginning of the Fifth Republic (1958). It appears clearly that the modifications that we conducted did not affect much the number of local positions (municipal, intercommunal, departmental, and regional councilors), as the red and green curves match very well (brown areas). This explains why the red line is barely visible in Figures 2.a—c and 2.f. The improvement lies elsewhere, in the quality and completeness of the term descriptions, rather than in their number. On the contrary, for the national positions (National Assembly, Senate, European Parliament), the plots show that we removed a fair number of spurious entries (red areas in Figures 2.d—e and 2.g) and added numerous missing ones (green areas in the same figures). This is not a surprise, as the secondary data sources that we integrated in addition to the RNE in order to constitute the BRÉF specifically focus on these types of position. Improving the coverage of the other types will require leveraging additional sources, a task which is currently ongoing.

Table 8 provides the overall numbers of terms of all types of position described in the BRÉF. The database is obviously dominated by the local positions, with the municipal councilors accounting for 89 % of the `Mandate` table. The table also exhibits the numbers of terms by gender, revealing an overall imbalance, with 63 % of terms held by males.

**Table 8.** Number of entries in data table `Mandate`, by type of position, date and gender.

| Type of position | Earliest term | Latest term | Total Number of terms | Number of terms by male representatives | Number of terms by female representatives |
|---|---|---|---|---|---|
| Department Councilor | 1977/03/14 | 2018/06/14 | 12,767 | 9,427 | 3,340 |
| Intercommunal Councilor | 1971/07/21 | 2018/06/30 | 125,226 | 86,702 | 38,524 |
| Municipal Councilor | 1952/10/17 | 2018/06/30 | 1,223,591 | 762,520 | 461,071 |
| Regional Councilor | 1986/03/16 | 2018/04/05 | 5,792 | 3,009 | 2,783 |
| Member of the National Assembly | 1986/03/16 | 2020/01/17 | 2,542 | 1,909 | 633 |
| President of the Republic | 1958/12/21 | 2017/05/06 | 13 | 13 | 0 |
| Member of the European Parliament | 1979/06/07 | 2019/05/25 | 893 | 594 | 299 |
| Senator | 1948/11/07 | 2019/09/30 | 3,272 | 2,913 | 359 |
| Total | 1952/10/17 | 2020/01/17 | 1,373,071 | 866,351 | 506,719 |

We now focus on field completeness. For the main fields of interest, Table 9 presents the numbers of entries that possess a value vs. those that are empty. The fields not mentioned here are very complete and do not require any particular discussion. In table `Individual`, all fields containing personal information are fully complete, except those related with the representatives' birthplace. This information was not available in the original RNE dataset, and only in some of the secondary sources. This can be an issue when matching the BRÉF individuals with those described in some other external databases, as names and birthdates are sometimes not enough to distinguish two persons. In table `Mandate`, the temporal boundaries are near-complete, especially the starting date. Note that the missing end dates correspond essentially to terms that are ongoing at the time the database was last updated, and are consequently not errors.

**Table 9.** Completeness of the main fields in our database.

| Table | Field | Filled | Missing | Completeness |
|---|---|---|---|---|
| Individual | `LastName` `FirstName` `Gender` `BirthDate` | 904,984 | 0 | 100.00 % |
| | `BirthMunicipality` | 1,523 | 903,461 | 0.17 % |
| | `BirthDepartment` | 1,425 | 903,559 | 0.16 % |
| | `BirthCountry` | 1,525 | 903,459 | 0.17% |
| Mandate | `MandateStartDate` | 1,373,071 | 0 | 100.00 % |
| | `MandateEndDate` | 798,725 | 574,346 | 58.17 % |
| | `MandateEndReason` | 795,829 | 577,242 | 57.96 % |
| | `PoliticalNuanceRNEId` | 1,373,069 | 2 | 100.00 % |
| | `ProfessionId` | 1,373,071 | 0 | 100.00 % |
| Function | `FunctionStartDate` | 272,713 | 324 | 99.88 % |
| | `FunctionEndDate` | 103,692 | 169,345 | 37.98 % |
| | `FunctionEndReason` | 86,081 | 186,956 | 31.53 % |
| | `FunctionType` | 273,037 | 0 | 100.00 % |

The end of term reason is missing for 2,896 of the completed terms, which will require some additional work. The concerned terms are mainly local ones, which are generally not as well described as national ones. The political orientation and profession of the representative at the time of the election are very well filled, but as free text: both of these fields require some normalization to be really useful. The situation is similar for table `Function`, although its completeness is lower because its fields are not as systematically filled in the RNE. In particular, the end of function reason misses from 17,611 entries, corresponding again mainly to local positions. The fields in the other tables (`PoliticalNuance`, `Profession`, `Area` and `Inclusion`) are very complete; these tables do not exhibit any particular situation worth mentioning.

**Table 10.** Reliability of the main fields in our database.

| Table | Field | Filled | Errors | Reliability |
|---|---|---:|---:|---:|
| `Individual` | `IndividualId` | 904,984 | 67 | 99.99 % |
| | `LastName` | 904,984 | 113 | 99.99 % |
| | `BirthDate` | 904,984 | 11 | > 99.99 % |
| `Mandate` | `MandateId` | 1,373,071 | 1105 | 99,92 % |
| | `MandateStartDate` | 1,373,071 | 605 | 99,96 % |
| | `MandateEndDate` | 798,725 | 114 | 99,99 % |
| | `MandateEndReason` | 795,829 | 1097 | 99,86 % |
| | `PoliticalNuanceRNEId` | 1,373,071 | 86 | 99,99 % |
| | `IndividualId` | 1,373,071 | 174 | 99,99 % |
| | `IndividualUsedName` | 1,373,071 | 96 | 99,99 % |
| | `AreaId` | 1,373,071 | 261 | 99,98 % |
| `Function` | `FunctionId` | 273,037 | 66 | 99,98 % |

After having assessed the completeness of our database, we now turn to the reliability of the data present in the BRÉF. There are two types of errors remaining in the database: on the one hand, those that we detected but have not corrected yet, and on the other hand, those of which we are not aware. It is difficult to estimate the latter, so we focus on the former: as a consequence, the statistics that we present and discuss here are only an approximation of the quality of our data. We used two different approaches to detect remaining errors in the BRÉF. The first approach is through the integration of additional secondary sources, which sometimes contradict the current data. We solve such conflicts manually, and possibly identify errors in the original data. This is an ongoing process, that will lead to a subsequent version of the BRÉF. Table 10 provides the numbers of errors detected for the main fields of interest, when ignoring all missing values. The fields (and data tables) not shown here correspond to fewer than 10 errors. Column *Reliability* shows the percentage of filled values that happen to be correct. In the case of `IndividualId` (table `Individual`), `MandateId` (table `Mandate`) and `FunctionId` (table `Function`), the errors correspond to duplicates or superfluous entries, which need to be removed. For the other fields, the errors correspond to incorrect values, which need to be fixed. Most of the errors remaining in the BRÉF concern duplicate mandates, and issues in mandate dates. Even for these fields, the proportion of errors relative to the number of filled entries is very low.

**Table 11.** Inconsistencies detected in our database.

| Consistency test | Mandates | Errors | Reliability |
|---|---|---|---|
| Overlapping mandates of the same type, for the same individual, and the same area. This query ignores municipal and intercommunal mandates. | 25,226 | 1 | > 99.99 % |
| Overlapping mandates of the same type, for the same individual, and the same area. This query focuses only on municipal and intercommunal mandates. | 1,347,844 | 1,031 | 99.92 % |
| Overlapping completed and ongoing mandates of the same type, for the same individual, independently of the area. | 798,724 | 1,019 | 99.87 % |
| Municipal mandates starting at the same date, for the same individual, but for different municipalities. | 1,223,079 | 206 | 99.98 % |
| Intercommunal mandates not matching municipal mandates with compatible dates. The intercommunal dates should be bounded by municipal dates. | 125,226 | 14,703 | 88.26 % |

The second approach that we use to detect errors consists in performing a set of consistency tests, based on geographical and legal constraints. Concretely, we have identified a set of rules that the data should enforce, and any exception is likely to correspond to an incorrect value. These errors are still present in this version of the BRÉF either because fixing them requires a manual intervention, or some data that we do not currently possess. As mentioned before, we aim at solving these problems in the next version of the BRÉF. Table 11 exhibits the results of these consistency tests, listing the numbers of detected errors. In order to help interpreting these numbers, the *Mandates* column provides an idea of the worst case, by indicating the number of mandates concerned by the test. Note that we also conducted other tests, but either they did not allow detecting any errors, or these errors are already corrected in the current version of the BRÉF. For instance, one such test looks for incompatible dates between a mandate and its associated function, and now finds none because we already tackled this issue completely. Due to this lack of detected errors, these tests are not presented in Table 11. However, they are available on our `BrefConversion` repository [18]. In Table 11, the first two tests are designed to detect duplicate mandates with slightly different dates; they differ in the mandate types they focus on. The third test aims at identifying situations were two consecutive mandates overlap, when they should not. The fourth corresponds to a legal impossibility, and allows detecting complete homonyms (same first name, last name, and birthdate) that are incorrectly considered as the same individual. Finally, the fifth test is used to find incorrect dates in intercommunal mandates. Overall, the number of errors is small relative to the number of concerned mandates. These errors tend to be concentrated among the local mandates, which are not as well documented as the national ones. Of course, this assessment of the BRÉF reliability is incomplete, as our tests are likely to cover only partially the space of all possible errors. We are confident they deal with the main types of erros, though, and provide a relatively accurate depiction of data quality in the BRÉF.

## 4. EXPERIMENTAL DESIGN, MATERIALS AND METHODS

This section describes how we elaborated the BRÉF. We first introduce our main data source (Section 4.1), before discussing the issues that these data exhibit (Section 4.2). We then explain the methods that we proposed and applied to tackle these issues and constitute the BRÉF (Section 4.3), as presented in Section 3.

### 4.1 Primary data source

As mentioned before, our primary source to build the BRÉF is the RNE. This database is managed by the *Bureau des Elections*[5] (BdE), a service under the French Ministry of the Interior, and authorized by decree [7]. Its main objective is to inform the legislative bodies, the government, and the population about elected officials and the terms they serve. It also aims to ensure that these terms comply with legislation, particularly regarding the limitation of concurrent terms [8] and gender parity [9]. The Ministry centralizes the data, but their entry is carried out at the prefecture level (i.e. department capital), mainly during elections.

The RNE is not directly accessible to the general public, probably because it contains sensitive information (such as the addresses of elected officials). For this reason, we do not know exactly what it contains, apart from the data made public by the Ministry. Indeed, as provided by the rule of communication of information [10], the Ministry is required to provide any person who requests it with a partial extraction of the RNE, from which personal information deemed sensitive is removed.

Until January 2019, this request was made to a prefecture, which then relayed it to the Ministry. It was possible to request two types of extractions: *cumulative* vs. *snapshot*. A cumulative extraction made on a given date contains the evolution of terms starting from the creation of the RNE in 2001 until the given date. In contrast, a snapshot extraction is limited to the list of elected officials in office *on the given date*. After January 2019, the Open Data law [11] changed the policy on the communication of administrative documents. The Ministry implemented a new mode of access by publishing snapshots of the RNE every three months on data.gouv.fr, the French government's Open Data website [12]. For the Ministry, such systematic and regular publication of the data is intended to replace the discretionary communication previously implemented. While this approach has the advantage of greatly facilitating access to the RNE, it also results in a reduction of the amount of available data, which can be a major problem for researchers. First, the datasets available on data.gouv.fr are not cumulative, and therefore miss historical depth. Second, several attributes describing the terms are omitted from the provided data: unique identifier of the elected officials, reasons for the end of term and function, and political nuance.

---
[5] French Bureau of Elections

Under the rule of communication of information, we requested the BdE as researchers, on the 9 February 2017, and obtained a cumulative extraction made on the 17 July 2018, which constitutes our main source in the creation of the BRÉF.

## 4.2 Detected problems

We analyzed thoroughly the data provided by the BdE, and detected 61 distinct types of issue, with varying levels of severity. We list them in Table 12 and summarize them in the following, but the interested reader will find a detailed description in our technical report [13] (in French).

**Table 12.** List of the types of issue identified in the raw RNE data.

| # | Issue | # | Issue |
|---|---|---|---|
| 1 | Inconsistent use of diacritics in names | 32 | Missing end of function dates |
| 2 | Consecutive space characters in names | 33 | Inconsistent start of function dates |
| 3 | Inconsistent use of punctuation in names | 34 | Inconsistent end of function dates |
| 4 | Inconsistent usual names | 35 | Inconsistent function time ranges |
| 5 | Missing first names | 36 | Impossibly short functions |
| 6 | Presence of typos | 37 | Time-overlapping functions |
| 7 | Inconsistent typographic case in names | 38 | Functions defined outside a term |
| 8 | Digits in canton names | 39 | Missing department IDs |
| 9 | Irrelevant information mixed with names | 40 | Incorrect department IDs |
| 10 | Articles occurring in toponyms | 41 | Non-unique department IDs |
| 11 | Inconsistent use of abbreviations | 42 | Non-unique constituency IDs |
| 12 | Missing municipality names | 43 | Non-unique canton IDs |
| 13 | Political nuance name for 'unknown' | 44 | Non-unique intercommunal IDs |
| 14 | Inconsistent political nuance names | 45 | Missing municipality IDs |
| 15 | Ambiguous political nuance names | 46 | Non-unique municipality IDs |
| 16 | Missing political nuance names | 47 | Inconsistent municipality IDs |
| 17 | Missing occupations | 48 | Several IDs for the same individual |
| 18 | Missing reason for end of term | 49 | Several individuals for the same ID |
| 19 | Missing reason for end of function | 50 | Contradictory entries |

| 20 | Inconsistent function names |
| 21 | Incomplete function names |
| 22 | Inconsistent names of reasons for end |
| 23 | Missing birthdates |
| 24 | Inconsistent birthdates |
| 25 | Missing start of term dates |
| 26 | Inconsistent end of term dates |
| 27 | Inconsistent term time ranges |
| 28 | Impossibly short terms |
| 29 | Time-overlapping terms |
| 30 | Term spanning several elections |
| 31 | Missing start of function dates |

| 51 | Incorrect constituencies |
| 52 | Missing individuals |
| 53 | Missing district councilors |
| 54 | Missing Corsica councilors |
| 55 | Missing Martinique/Guyane councilors |
| 56 | Missing territorial councilors |
| 57 | Missing New Caledonia councilors |
| 58 | Incomplete temporal coverage |
| 59 | Missing terms |
| 60 | Spurious terms |
| 61 | Incorrect number of elected positions |

**Form, validity & completeness.** The problems of form concern the way information is represented in the data tables, those of validity pertain to the very nature of this information, and the notion of completeness refers to the unfilled values. These problems appear quite frequently in the RNE database, and although they may seem trivial, they nevertheless lead to sometimes significant complications during automatic data comparison. For this reason, it is necessary to resolve them first.

**Data consistency.** The problems of consistency between values mainly affect dates. A good number of them can be at least detected, if not resolved, by automatic means, by cross-referencing the information contained in the database and comparing it with external resources such as those described later (election dates, number of seats per institution, etc.).

**Bijective relations.** The expected bijective nature of the relations between entities and identifiers is lacking in many of the relevant fields. Again, the use of external resources should help resolve part of the problem, particularly those concerning administrative areas. Resolving the issues with identifiers of elected officials, who constitute the central entity of the database, is more difficult.

**Reliability of terms.** The fact that on the one hand, there are missing terms in the RNE, and on the other hand, many terms are erroneous, constitutes a critical and major difficulty. Addressing this requires the use of secondary sources that can be cross-referenced at least partially with the RNE, if only to assess its levels of incompleteness and error.

We identified three potential causes to these issues: 1) the data entry process at the prefecture level; 2) the structure of the Ministry's internal database; and 3) the extraction process that produced the cumulative database. The data entry process appears inconsistent across operators, leading to variable interpretations and input methods. For instance, some operators visibly misinterpret fields related to the end of a term, resulting in incorrect dates. The graphical interface used might contribute to errors through features like autocomplete, which may assign terms to the wrong individuals or incite to create duplicate entries. The lack of validation checks during data entry, possibly to avoid discouraging operators, also plays a role. The RNE database structure seems not designed for longitudinal data storage, causing issues with area identifiers and hindering the representation of their evolving characteristics. Finally, the extraction process itself might introduce errors, such as inconsistent birthdates for elected officials.

## 4.3 Data processing

We have produced two major versions of the database, as detailed in Table 13. The first major version takes the form of a single large CSV file, whereas the second major version is a proper relational database, whose structure is described in Section 3. The whole database elaboration process includes a number of operations, that can be separated in three phases: first, we corrected a number of issues directly in the RNE data; second, we leveraged several secondary sources to get missing information and correct data errors; and third, we populated the PostgreSQL database and corrected a number of small minor remaining issues detected during this step. The scripts implementing this processing are available on two GitHub repositories: `BrefInit` for everything related to the first version [17], and `BrefConversion` for the second version [18]. In addition, `BrefInit` also contains all the raw data required to produce the first version of the BRÉF, including the main and secondary sources.

**Table 13.** Versions of the BRÉF leading to the database presented in this article.

| Version | Date | Description |
|---------|------|-------------|
| 1.0.0 | 2020/06/24 | Analysis and cleaning of RNE database, and integration of the data coming from the National Assembly [15], Senate [16], and European Parliament [17] databases. All raw datasets are available on `BrefInit`, our first GitHub repository [17]. The result is a large CSV file with roughly the same structure as the original RNE data. |
| 1.0.1 | 2020/07/08 | Correction of minor errors: missing first names, incorrect IDs and last names, departments of some EPCI. |
| 1.0.2 | 2020/07/18 | Correction of a few minor errors related to EPCI IDs and departments. This version of the database is described in our technical report [14]. |
| 1.1.0 | 2020/09/17 | In the process of switching to a PostgreSQL system, we detected a number of issues through the implementation of the constraints described in Section 3.3 (inconsistencies between terms and functions, problems with area IDs, etc.). This version is the result of their correction. |
| 2.0.0 | 2020/11/27 | Creation of the database with the architecture described in Section 3, and migration of data from the single table from v1.1.0 to a PostgreSQL system. |
| 2.0.1 | 2020/11/27 | Correction of some issues, concerning profession, found during the integration of version 2.0.0; creation of field `CodeTerritoire` |

| | | for overseas territories. |
|---|---|---|
| 2.0.2 | 2021/01/06 | Corrections regarding other term duplicates, more specifically, those differing only by their end motive. |
| 2.1.0 | 2021/06/09 | Additional adjustments aiming at preparing the database for the future integration of data coming from new alternative sources. |
| 2.2.0 | 2024/08/29 | This version was created for the data paper. It corresponds to v2.1.0 with a few additional constraints, and table and field names translated to English. |
| 2.2.1 | 2025/01/15 | Correction of several errors in the field names of v2.2.0. This is the version presented in this article [7]. |

**Correction of the RNE data.** The process involves 13 types of modifications, which are detailed comprehensively in our technical report [13]: here we only give a very brief summary of the operations conducted on the RNE data. We first normalized the names, by consistently handling diacritics, punctuation, spaces, case, abbreviations, and numbering. Second, we automatically detected homonyms among individuals, and manually resolved situations corresponding to duplicates. Third, we automatically detected inconsistencies in names, genders, dates, and manually solved them using online resources. Fourth, we defined *ad hoc* rules to conduct a number of automatic corrections regarding the usage of maiden names, the uniformization of political nuance names, the completion of missing area IDs. Fifth, we added some new fields, to account for missing information, and also to mark the database entries undergoing some modification during this correction process. Sixth, we merged redundant term and function entries. Seventh, we aligned incorrect function dates with term dates. Eighth, we aligned incorrect term dates with election dates. Ninth, we suppressed very short terms (a few days), which typically correspond to mishandled situations of a substitute filling in for the elected person. Tenth, we merged terms of the same individual (and type) exhibiting temporal overlap. Eleventh, we split terms spreading over several elections. Twelfth, we fixed incompatible dates in terms involving different consecutive persons on the same unique position. Thirteenth and finally, we detected and fixed end of term reasons that were incompatible with the term information. This process is implemented in a collection of R scripts, available on GitHub repository `BrefInit` [16].

**Integration of the secondary sources.** Many of the issues listed in Section 4.2 could be corrected through the operations described in the previous paragraph, because they could be detected by cross-comparison with other entries of the RNE, and fixed automatically via some custom programs or manually based on online information. However, it was not the case of all the issues: for some of them, the missing information or incorrect value could not be inferred or reconstructed based on the rest of the data. This time, the operation required to leverage other data sources. This is what we did, by using public open data coming from the French National Assembly [14], the French Senate [15], and the official Website of the European Parliament [16]. All three sources are cumulative, in the sense that they give access to all the individuals elected in the concerned institutions after a certain date. The general process was the same for all three sources: we first matched the individual from the RNE and the considered database, based on their personal information. Then, we matched their terms and functions, based on their natures, dates and constituencies. Finally, we integrated the external information into our database. We adopted a semi-automatic approach to verify cases where the secondary source brought new information (i.e. completely absent from the RNE), or was inconsistent with the RNE data. Indeed, although their data turned out to be more reliable than the RNE's, these secondary sources were not devoid of any problems themselves, which is why we carefully integrated their information into our database. This process is implemented as of R scripts, which are publicly available on the previously mentioned `BrefInit` repository [17].

**Switch to PostgreSQL.** At this stage, the database reached its first major version, taking the form of a single table containing all the information. This format is seemingly easier to use by people without any database expertise, but it is harder to maintain in the long term due to information duplication and lack of structure. For these reasons, we decided to switch to a proper relational database. Using a PostgreSQL system, we created the 7 tables described in Section 3, according to the model pictured by Fig.1. We imported the first version of our data into a temporary table, and designed a collection of PL/pgSQL scripts to transfer them from this temporary table into the appropriate database tables, in accordance with the integrity constraints. This integration allowed us to detect additional issues in the data, especially strict term duplications, which we deleted. After integration, we handled additional issues manually, in particular the representation of overseas territories and French ex-colonies, town and canton duplications, and area inclusion problems. Some function inconsistencies revealed that 136 terms were missing, which we inserted to the database. The resulting database [6] is the version presented in this article. The PL/pgSQL scripts are publicly available on GitHub repository `BrefConversion` [18].

# LIMITATIONS

The BRÉF encompasses a wide range of terms, covering the whole French Fifth Republic, from 1958 (its creation) to 2018 (last cumulative RNE extraction). For this purpose, it integrates data coming from four different sources, characterized by varying levels of completeness and relevance. This heterogeneity makes it difficult to obtain a perfectly complete and reliable database: the BRÉF still contains a few spurious terms, and lacks a number of local terms. Moreover, some terms are incompletely described. The reliability is very good on the national positions, though, due to the cross-checking with secondary sources focusing on these specific types. Experiments were conducted on this version of the BRÉF [2], and the obtained results are consistent with the literature in Political Science and demonstrate the interest of this database.

In order to solve the current limitations of the BRÉF, we are currently integrating another secondary source. According to our estimation, this will allow retrieving a large part of the missing terms, and correct certain errors currently present in the database. In parallel, we are setting up a process aiming at integrating to the BRÉF the RNE snapshots published quarterly by the French Ministry of the Interior, in order to keep the database up-to-date.

# ETHICS STATEMENT

The BRÉF compiles personal data from the RNE and from the open databases of the French National Assembly, the French Senate, and the European Parliament. These data originally come, on the one hand, from the registration of electoral candidates and the regular updates of the lists of elected representatives conducted in the French prefectures and, on the other hand, from the administrative registration of the elected representatives at the parliamentary assemblies. The RNE has been set up for the administrative control of candidates and elected representatives, but also "with a view to informing the public authorities and citizens", according to article 4 of a 2014 decree [19]. This decree does not specify whether the consent of election candidates is obtained when they submit their candidatures. However, its article 8 explicitly specifies that any person may be given access to data and information on request, except for their mail address, telephone number and e-mail address.

At this stage, the BRÉF does not contain any questionnaire survey data. As the database is biographical in nature, the data is not anonymized. The declared profession and political affiliation of elected representatives are the fields that give the database its sociological value. These data may be considered sensitive, in particular the latter fields, in accordance with the GDPR. However, this information is already freely accessible in the lists of candidates for local and national elections, which are published in France under open licenses [12] [20].

We have ensured that our research practices fully comply with the ethical requirements for publication in the Data in Brief journal.

# CRediT AUTHOR STATEMENT

Noémie Févrat: Data curation, Conceptualization, Methodology, Investigation, Validation.

Vincent Labatut: Data curation, Conceptualization, Methodology, Software, Writing, Supervision.

Émilie Volpi: Data curation, Conceptualization, Methodology, Software, Writing.

Guillaume Marrel: Resources, Conceptualization, Methodology, Investigation, Validation, Writing, Supervision.

# ACKNOWLEDGEMENTS

Funding: This work was supported by the research federation Agorantic (FR 3621), through the funding of Noémie Fevrat's PhD. and of project BRÉF2.

Data: we thank the Office of Elections form the French Ministry of the Interior for providing us with a part of the data included in the BRÉF.

Non-author contributors: Cédric Richier (Agorantic research federation) participated in data curation and programming. Youba Aguanana (Avignon Université) contributed to data curation.

# DECLARATION OF COMPETING INTERESTS

The authors declare that they have no known competing financial interests or personal relationships that could have appeared to influence the work reported in this paper.